\documentclass[twocolumn,floats,groupedaddress,prb,aps]{revtex4}

\usepackage[T1]{fontenc}
\usepackage{lmodern}
\usepackage[english]{babel}

\usepackage{mathrsfs}
\usepackage{amsfonts}
\usepackage{amssymb}
\usepackage{graphicx}
\usepackage{dcolumn} 
\usepackage{amsmath}
\usepackage{bm}      
\usepackage{graphicx}
\usepackage{color}
\usepackage{epsfig,psfrag,subfigure,subeqnarray}
\def\lan{\langle}
\def\ran{\rangle}
\def\va{\varepsilon}
\def\dag{\dagger}

\def\vk{{\bf k}}

\def\vp{{\bf p}}
\def\vq{{\bf q}}
\def\vr{{\bf r}}

\def\v0{{\bf 0}}

\def\vS{{\bf S}}
\def\vL{{\bf L}}
\def\vJ{{\bf J}}

\newcommand{\bd}{\begin{equation}}
\newcommand{\ed}{\end{equation}}
\newcommand{\be}{\begin{equation}}
\newcommand{\ee}{\end{equation}}
\newcommand{\bt}{\begin{split}}
\newcommand{\et}{\end{split}}

\newcommand{\bn}{\begin{align}}
\newcommand{\en}{\end{align}}
\newcommand{\bea}{\begin{eqnarray}}
\newcommand{\eea}{\end{eqnarray}}
\newcommand{\ba}{\begin{array}}
\newcommand{\ea}{\end{array}}
\newcommand{\nn}{\nonumber}

\DeclareMathAlphabet\mathbfcal{OMS}{cmsy}{b}{n}

\begin{document}
\title{Exploring the change of  semiconductor hole mass under Coulomb scattering}

\author{Shiue-Yuan Shiau}
\affiliation{Physics Division, National Center for Theoretical Sciences, Hsinchu, 30013, Taiwan}
\author{Monique Combescot}
\affiliation{Institut des NanoSciences de Paris, Sorbonne Universit\'e, CNRS, 4 place Jussieu, 75005 Paris}

\begin{abstract}
Semiconductor valence holes are known to have heavy and light effective masses; but the consequence of this mass difference on Coulomb scatterings has been considered intractable and thus ignored up to now. The reason is that the heavy/light index is quantized along the hole momentum that changes in a Coulomb scattering; so, a heavy hole can turn light, depending on the scattering angle. This mass change has never been taken into account in many-body problems, and a single ``average'' hole mass has been used instead. In order to study the missed consequences of this crude approximation, the first necessary step is to determine the Coulomb scatterings with valence holes in a precise way. We here derive these scatterings from scratch, starting from the threefold valence-electron spatial level, all the way through the spin-orbit splitting, the Kohn-Luttinger effective Hamiltonian, its spherical approximation, and the phase factors that appear when turning from valence electron to hole operators, that is, all the points of semiconductor physics that render valence holes so different from a na\"{i}ve positive charge.   
\end{abstract}
\date{\today}
\maketitle

Semiconductors continue to be of high technological interest in many industrial sectors\cite{Gmachl,Sang}. The prime advantage of these materials is their complex, yet controllable, band structure\cite{Esaki1970,BastardIeee,Capasso1987}, with energy levels that produce desirable effects for electronic and optoelectronic applications. 

Among the band structure complexities, one  is linked to the upper valence states. These states come from a degenerate spatial level which is partly split by the spin-orbit interaction. This interaction, which  in the case of atomic electrons\cite{Sakurai} takes the  familiar $\hat{\vL}\cdot\hat{\vS}$ form in terms of their orbital angular momentum $\hat{\vL}=\vr\times\hat{\vp}$, demands a totally different approach for electrons in periodic Bloch states because the potential felt by these electrons is not spherical, so that the spin-orbit interaction does not read in terms of $\vr\times\hat{\vp}$ in the case of  semiconductor electrons. For this reason, the crystal symmetry and periodicity are commonly tackled through the group theory\cite{Falicov,Bir,Ivchenko}. The disadvantage of this approach is to shade the physical origin of the splitting when spin is added, \textit{i.e.}, when the ``simple'' group is transformed into the ``double'' group. This is why we have recently proposed a totally different approach\cite{MonicPRB2019}. Through a $\textbf{k}\cdot\textbf{p}$ perturbative procedure\cite{Voonbook,Fishman},  we have recovered that the  dispersion relation  of the upper spin-orbit valence electrons  has two warped surfaces\cite{Dresselhaus,LK,luttinger}, pinned on the crystal axes. They  are commonly approximated as two spheres associated with heavy and light masses\cite{Bastardbook,Cardona}. Heavy and light holes have been experimentally evidenced  in bulk GaAs by using optically pumped NMR\cite{Ramas,Dustin}, despite their small energy difference.

Getting rid of the warping, which removes the crystal axes from the electron dispersion relation, is already a big simplification toward handling valence electrons. Still, these axes do not completely disappear from the problem due to Coulomb interaction. Here enters a ``miracle'', that most people take as granted without well-established derivation: the Coulomb scatterings for \textit{intra}band processes between Bloch-state electrons in the conduction and valence bands,  have the same value as the one for electrons in vacuum --- within a dielectric constant reduction that comes from \textit{inter}band processes\cite{Monicbook}. However, these scatterings, derived in the Bloch basis with states labeled as $\mu =(x,y,z)$ along the crystal axes, have the  $1/q^2$ dependence \textit{exclusively} for processes that are diagonal in $\mu$. It is in this tricky way that the crystal axes enter the Coulomb potential in semiconductors. 

By contrast, the crystal axes do not appear in the heavy and light valence states because they are eigenstates of a \textit{spherical} Hamiltonian. Thus, only one axis  remains to classify the fourfold valence subspace: the $\vk$ wave vector of the electron. Indeed, the heavy and light valence electrons are  distinguished by a fourfold index $\mathcal{J}=(\pm3/2,\pm1/2)$ with quantization axis along \textbf{k}.  As a direct consequence, a valence electron in a heavy state when its wave vector is $\vk$, does not necessarily stay heavy when its wave vector changes.

The fact that the Coulomb interaction is not diagonal between valence electron eigenstates has a dramatic consequence on many-body effects involving these electrons: because the electron wave vector $\vk$ changes in a Coulomb process, so changes the valence electron state along with its mass. The heavy-light hole mass difference has been deemed  intractable in many-body problems. This is why, up to now, all many-body effects in bulk semiconductors have been derived by adopting
 a single hole mass, obtained from the average of the inverse heavy-hole and light-hole masses\cite{MC1972,MC1972b,Brinkman1972}. While such average seems reasonable when dealing with electron-hole binding, it is hard to accept for problems in which the pair center of mass plays a role. Up to now, there is no satisfactory treatment to this mass problem. Note that the situation is  different in quantum wells because the heavy-hole and light-hole energies are split by confinement\cite{Laruelle}; so, in narrow quantum wells, we can just keep Coulomb processes within the heavy-hole subspace.
 
Our motivation for deriving the Coulomb scatterings between heavy and light holes is to reconsider the validity of using a single hole mass in many-body problems, and to provide some mathematical support to the widely-used average hole mass value. The validity of using a single mass most probably depends on the problem at hand. A first question surely is related to the exciton\cite{Monicbook}: does the Coulomb coupling between heavy and light holes split the exciton degeneracy obtained by using a single average hole mass? In view of the narrow exciton line, such a  splitting should be observable through optical experiments in bulk samples\cite{Ramas,Dustin}.
 
The very first step to investigate the consequences of using an average hole mass in bulk many-body effects---which still is an open problem---is to control how a heavy valence electron stays heavy or turns light under Coulomb scatterings. This requires to go back to the microscopic form of the Coulomb interaction in semiconductors\cite{Monicbook}, originally written in terms of valence electrons in Bloch states $\mu$. By rewriting the creation operators of these  electrons in terms of heavy-mass and light-mass states, we see that the Coulomb scatterings, diagonal in $\mu$, do not stay diagonal for heavy and light valence electrons. This is the fundamental reason why the valence hole mass can change in a Coulomb process. Quantities of physical interest are the probabilities for mass change, heavy to heavy (HH) or heavy to light (LH); they depend on the angle $\theta_{\vk'\vk}$ between the outgoing and incoming hole wave vectors $\vk'$ and $\vk$, in an amazingly simple way:
\bea
1-\frac{3}{4}\sin^2\theta_{\vk'\vk}=P_{HH}=1-P_{LH}
\nonumber
\\
=P_{LL}=1-P_{HL}\hspace{0.1cm}\label{eq1}
\eea
The fact that there are two heavy states and two light states, which restores  time reversal symmetry, plays a crucial role in the simplicity of these results. Note that  $P_{LH}=P_{HL}$ is easy to expect  physically. It implies the less obvious relation $P_{HH}=P_{LL}$, which mathematically follows  from  the probability sum to be equal to 1.
 Unfortunately, the microscopic scatterings between  heavy and light states, which are the relevant quantities to derive many-body effects, are not as simple as these probabilities.
 
\begin{figure}[t]
\begin{center}
\includegraphics[trim=1cm 3cm 1cm 4cm,clip,width=3.5in]{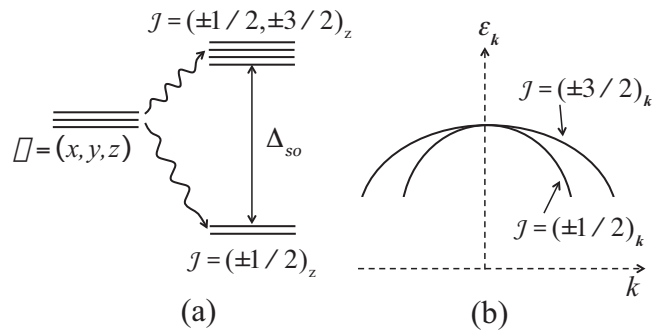}
\end{center}
\vspace{-0.4cm}
\caption{(a) When spin is added, the spin-orbit interaction splits the sixfold level originating from the threefold spatial level $\mu=(x,y,z)$ into a fourfold level $\mathcal{J}_z $ with $\mathcal{J}=(\pm3/2,\pm1/2)$ and a twofold level with $\mathcal{J}=\pm1/2$. (b) Within the spherical approximation, the eigenstates for $\vk$ electron in the fourfold level are made of two heavy states $\mathcal{J}=(\pm3/2)_\vk$ and two light states $\mathcal{J}=(\pm1/2)_\vk$, quantized along the $\vk$ direction.   }
\label{appCoulomb_HL:fig2}
\end{figure}

\textbf{We here show} that the Coulomb scatterings involving heavy and light valence electrons read as the Coulomb scattering for conduction electrons, $ 4\pi e^2/\epsilon_{sc}L^3 q^2$, multiplied by a complex factor ${}_{\vk'}\lan\mathcal{J}'|\mathcal{J}\ran_{\vk}$ that corresponds to the scalar product of the incoming and outgoing valence electron eigenstates for $\vk$ and $\vk'$ wave vectors. These $\mathcal{J}$ indices, quantized along the electron wave vector,  have been commonly taken as $(\pm 3/2)$ for heavy electrons and $(\pm 1/2)$ for light electrons (see Fig.~\ref{appCoulomb_HL:fig2}(b)), due to similarity with their atomic counterparts. The ${}_{\vk'}\lan\mathcal{J}'|\mathcal{J}\ran_{\vk}$ factor reduces to $\delta_{\mathcal{J}',\mathcal{J}}$ for $\vk'$  parallel to $\vk$ and to $\delta_{\mathcal{J}',-\mathcal{J}}$ for $\vk'$ antiparallel to $\vk$: for such wave vectors, the valence electrons keep their mass. Otherwise, there is a non-zero probability for a mass change.

To the best of our knowledge, we have not found any derivation of the microscopic Coulomb potential given in Eq.~(\ref{app:hlhole40}). By using these scatterings, we recover the  compact result given in Eq.~(\ref{eq1}). These probabilities have been previously quoted\cite{MC1972,MC1972b} but without any derivation\cite{footnote1}. 

The paper is organized as follows:

 $\bullet$ In Section \ref{sec1}, we go all the way from valence electrons with spin $s_z$ in a threefold degenerate spatial level with Bloch states labeled as $\mu=(x,y,z)$ along the crystal axes, to the spin-orbit splitting, the dispersion relations for finite electron wave vector  $\vk$ and the corresponding eigenstates obtained from the  spherical approximation of the  Kohn-Luttinger Hamiltonian, from which heavy and light valence electrons are obtained. We  pay a special attention to the quantum indices that label these different valence states.

  $\bullet$ In Section \ref{sec2}, we consider  Coulomb interaction. It is diagonal in the Bloch-state basis, and stays diagonal in the spin-orbit basis when the fourfold quantum index is quantized along  a crystal axis. The Coulomb interaction becomes nondiagonal when the spherical spin-orbit eigenstate basis is introduced because the quantization axis is not along a fixed crystal axis, but along the electron wave vector $\vk$. 

  $\bullet$ In Section \ref{sec3}, we turn from valence-electron creation operator to hole destruction operator, taking into account  the phase factors induced by the degeneracies of the spatial and spin states. We ultimately obtain the microscopic Coulomb potential (\ref{app:hlhole40}) between an electron and a heavy or light hole.

 $\bullet$ In Section \ref{sec4}, we combine these fundamental  results of semiconductor physics, to obtain the transition probabilities between heavy and light holes induced by the Coulomb scatterings, given in Eq.~(\ref{eq1}).

 $\bullet$ We then conclude.

Before going further, we wish to comment on using notations borrowed from atomic physics when dealing with semiconductors. Because semiconductor electrons are in a periodic potential, they do not have the orbital angular momentum $\hat{\vL}=\vr\times \hat{\vp}$  specific to systems with spherical symmetry\cite{VAKu}, with state parity depending on orbital quantum number $\ell$ as $(-1)^\ell$. A na\"{i}ve extension of atomic notations to periodic systems can lead to incorrect results, as for example  seen from the fact that the lowest spatial conduction level in a cubic semiconductor is called S, while all conduction levels are odd in parity; in the same way, the highest spatial valence level is called P, while all valence levels are even. So, atomic notations are valid to label level degeneracy---here nondegenerate and threefold---but are not valid for state parity, which plays a crucial role in many electronic transition problems.

Actually, it is possible to define a vector operator $\hat{\mathbfcal{L}}$ from the potential $\mathcal{V}(\vr)$ felt by the electrons, 
\be
\nabla \mathcal{V}(\vr)\times \hat{\vp}\equiv\tilde{\lambda}\,\hat{\mathbfcal{L}}
\ee
that reduces to $\hat{\vL}$ when $\mathcal{V}(\vr)$ depends on $|\vr|$ only, as for spherical symmetry, but differs from $\hat{\vL}$ otherwise. For cubic symmetry, the commutation relations of the $\hat{\mathbfcal{L}}$ components are the same as those for $(\hat{L}_x,\hat{L}_y,\hat{L}_z)$, provided we align the arbitrary Cartesian axes $({\bf x},{\bf y},{\bf z})$ of the spherical system along the cubic crystal axes. When spin is introduced, the matrix  representations of the resulting $\hat{\mathbfcal{J}}=\hat{\mathbfcal{L}}+\hat{\vS}$ operator are identical to those of the $\hat{\vJ}=\hat{\vL}+\hat{\vS}$ operator, provided we again align the axes for $(\hat{J}_x,\hat{J}_y,\hat{J}_z)$ along the cubic crystal axes.

Because of identical commutation relations for $(\hat{\mathbfcal{L}},\hat{\vL})$ or $(\hat{\mathbfcal{J}},\hat{\vJ})$ components, we could be led to call  angular momentum $\hat{\mathbfcal{J}}$ as well as $\hat{\mathbfcal{L}}$. However, being aware of the lack of spherical symmetry that $\hat{\mathbfcal{L}}$  has and the aforementioned parity problem, we here prefer to securely derive all results from scratch, and only at the end draw connections with the angular momentum formalism. This is the price to pay for not using group theory, as commonly done in semiconductor physics.

\section{Valence electrons\label{sec1}}

 Electrons in the periodic potential of a semiconductor crystal are characterized by a plane wave  with wave vector $\vk$, modulated by a Bloch function $u_{n,\vk}(\vr)$ that has the crystal periodicity. The resulting electron wave function in a sample volume $L^3$ reads
 \be
 \label{2}
 \lan \vr|n,\vk\ran= \frac{e^{i\vk\cdot \vr}}{L^{3/2}}  u_{n,\vk}(\vr)
 \ee
 with a band index $n=c$ for nondegenerate conduction electrons and $n=(v,\mu)$ for valence electrons in a threefold spatial level with states labeled as $\mu=(x,y,z)$ along the cubic crystal axes. In addition to  spatial degeneracy,  electrons also have a spin degeneracy.

\subsection{Zero wave vector }

\noindent $\bullet$ As visualized in Fig.~\ref{appCoulomb_HL:fig2}(a), the spin-orbit interaction splits\cite{MonicPRB2019} the threefold spatial level of valence electron states $|v,\vk={\bf0};\mu\ran \otimes |s\ran_z$ with spin quantized along the crystal axis $\textbf{z}$, into a fourfold level
\be
|v,{\bf0}\ran\otimes |\mathcal{J}\ran_z\qquad\textrm{with}\qquad  \mathcal{J}=(\pm 3/2,\pm 1/2)\label{app:hlhole1}
\ee 
and a twofold level
\be
|v,{\bf0}\ran\otimes |\mathcal{J}\ran_z\qquad\textrm{with}\qquad  \mathcal{J}=(\pm 1/2)\hspace{1cm}\label{app:hlhole2}
\ee

\noindent $\bullet$ These spin-orbit eigenstates are labeled in the same way as the spin-orbit eigenstates for atoms, due to their similar forms. Indeed, if the arbitrary axes of a spherically symmetric system are taken along the cubic semiconductor axes, they are both related to the $|v,{\bf0};\mu\ran\otimes|s\ran
_z$ states through
\begin{eqnarray}
\!\!\!\!\!\!\!|v,{\bf0}\ran\otimes\left|\pm\frac 3 2\right\ran_z\!\!\!&=&\!\!|v,{\bf0};\pm1\ran_z\otimes \left|\pm\frac 1 2\right\ran_z
\nn
\\
\!\!\!\!\!\!\!|v,{\bf0}\ran\otimes\left|\pm\frac 1 2\right\ran_z\!\!\!&=&\!\!\frac{1}{\sqrt{3}}\bigg(|v,{\bf0};\pm1\ran_z\otimes \left|\mp\frac 1 2\right\ran_z
\label{5}
\\
&&+\sqrt{2}\,|v,{\bf0};0\ran_z\otimes \left|\pm\frac 1 2\right\ran_z \bigg)
\nonumber
\end{eqnarray}
for the fourfold level and 
\bea
|v,{\bf0}\ran\otimes\left|\pm\frac 1 2\right\ran_z\!=\pm\frac{1}{\sqrt{3}}\bigg(\sqrt{2}|v,{\bf0};\pm1\ran_z\otimes \left|\mp\frac 1 2\right\ran_z\nn
\\
-|v,{\bf0};0\ran_z\otimes \left|\pm\frac 1 2\right\ran_z \bigg)
\label{app:hlhole5}
\eea 
for the twofold level, the spatial parts of these states being related to the $|v,{\bf0};\mu\ran$ Bloch states by
\begin{subeqnarray}\label{app:hlhole6_0}
|v,{\bf0};\pm1\ran_z&=&\frac{\mp i|v,{\bf0};x\ran+|v,{\bf0};y\ran}{\sqrt{2}}\slabel{app:hlhole6}\\
|v,{\bf0};0\ran_z&=&i|v,{\bf0};z\ran\slabel{app:hlhole7}
\end{subeqnarray}
with phase factors taken along the Landau-Lifshitz's phase factors for spherical harmonics\cite{Landau}.

\subsection{Finite wave vector $\vk$}

\noindent $\bullet$ Away from the $\vk=\bf0$ point, the $\vk$ dependence  of the spin-orbit eigenstates $|v,\vk\ran\otimes |\mathcal{J}\ran_z$ is derived\cite{MCPRB2020} through the $\vk\cdot\vp$ coupling of the  $|v,{\bf0}\ran\otimes |\mathcal{J}\ran_z$  valence states  to the two lowest spatial levels of the conduction band, $|c,{\bf0}\ran$ and $|c,{\bf0};\mu\ran$, that respectively are nondegenerate and threefold.

These couplings lead to a $4\times4$ matrix in the fourfold spin-orbit subspace that can be written in a compact form\cite{MCSS2021}  in terms of $\hat{\mathbfcal{J}}=(\hat{\mathcal{J}}_x,\hat{\mathcal{J}}_y,\hat{\mathcal{J}}_z)$, which have the same values and thus the same commutation relations as the angular momentum $\hat{\vJ}$ for spherical systems. The resulting effective Hamiltonian reads
\bea
\hat{H}^{(\rm warp)}_\vk=\frac{1}{2m_0}\Big(\alpha_1  k^2 \hat{\mathbfcal{J}}^2
\!\!&{+}&\!\!\alpha_2\left(\vk\cdot \hat{\mathbfcal{J}}\right)^2\label{app:hlhole9}
\\
\!\!&{+}&\!\!\alpha_3\left(k_x^2 \hat{\mathcal{J}}_x^2+k_y^2 \hat{\mathcal{J}}_y^2+k_z^2 \hat{\mathcal{J}}_z^2\right)\Big)
\nonumber
\eea
the matrix for the $\hat{\mathbfcal{J}}$ component along $\textbf{z}$  in the $|v,{\bf0}\ran\otimes |\mathcal{J}\ran_z$ basis made of the $\hat{\mathcal{J}}_z$ eigenstates, reading, as expected for eigenstates, as
\be
\hat{\mathcal{J}}_z=\hbar\left(\begin{matrix}
\frac{3}{2} & 0 & 0 & 0 \\
0& \frac{1}{2} & 0 & 0 \\
0 & 0 & -\frac{1}{2} & 0 \\ 
0 & 0 & 0 & -\frac{3}{2}
\end{matrix}\right)_{z}\label{app:hlhole10}
\ee
The $\hat{H}^{(\rm warp)}_\vk$ Hamiltonian has the  form of the Kohn-Luttinger's Hamiltonian\cite{LK,luttinger} when the twofold split-off band is dropped.

The $\alpha_n$ coefficients in Eq.~(\ref{app:hlhole9}) are related\cite{MCPRB2020} to the microscopic couplings $\zeta_1$ and $\zeta_3$ of the $|v,{\bf0};\mu\ran$ states to the $|c,{\bf0}\ran$ state and $|c,{\bf0};\mu'\ran$ states,  as
\bea
\zeta_1&=&\frac{2}{m_0E_g}\big|\lan c,{\bf0}|\hat{p}_x|v,{\bf0};x\ran\big|^2
\\
 \zeta_3&=&\frac{2}{m_0(E_g+\Delta_c)}\big|\lan c,{\bf0};y|\hat{p}_x|v,{\bf0};z\ran\big|^2
\eea
where $m_0$ is the free electron mass, $E_g$ is the band gap, and $\Delta_c$ is the energy difference between the $|c,{\bf0}\ran$ and $|c,{\bf0};\mu\ran$ conduction levels. The $\alpha_n$ coefficients precisely read
\be
\alpha_1=\frac{4-3\zeta_1-\zeta_3}{15}\qquad\alpha_2=\frac{\zeta_1+\zeta_3}{3}\qquad \alpha_3=-\frac{2\zeta_3}{3}
\ee

\noindent $\bullet$ It is of interest to note that the $\alpha_3$ term of the $\hat{H}^{(\rm warp)}_\vk$ Hamiltonian, also reads
 \bea\label{eq13}
k_x^2 \hat{\mathcal{J}}_x^2+k_y^2 \hat{\mathcal{J}}_y^2+k_z^2 \hat{\mathcal{J}}_z^2
=\left(\vk\cdot \hat{\mathbfcal{J}}\right)^2
\,\,\,\,\,\,\,\,\,\,\,\,\,\,\,\,\,\,\,\,\,\,\,\,\,\,\,\,\,\,\,\,\,\,\,\,\,\,\,\,\,\,\,\,\,\,\,\,\,\,\,\,\,\,\,
\\
-\left\{k_xk_y\Big[\hat{\mathcal{J}}_x,\hat{\mathcal{J}}_y\Big]_++k_yk_z\Big[\hat{\mathcal{J}}_y,\hat{\mathcal{J}}_z\Big]_++k_zk_x\Big[\hat{\mathcal{J}}_z,\hat{\mathcal{J}}_x\Big]_+\right\}\nn
 \eea
 with $\big[\hat{\mathcal{J}}_x,\hat{\mathcal{J}}_y\big]_+=\hat{\mathcal{J}}_x\hat{\mathcal{J}}_y+\hat{\mathcal{J}}_y\hat{\mathcal{J}}_x$.  When inserting this result into $\hat{H}^{(\rm warp)}_\vk$, we find that the above curly bracket is responsible for the warped contribution to the eigenenergies, in $\left(k_x^2k_y^2+k_y^2k_z^2+k_z^2k_x^2\right)$, while the other terms of $\hat{H}^{(\rm warp)}_\vk$ produce spherical contributions in $k^2$.
 
 \subsection{Heavy and light valence electrons }
 
The warping is commonly neglected by approximating the $\hat{H}^{(\rm warp)}_\vk$ Hamiltonian by a spherical Hamiltonian
\be
\hat{H}^{(\rm sph)}_\vk=\frac{1}{2m_0}\Big(\beta_1  k^2 \hat{\mathbfcal{J}}^2+\beta_2\left(\vk\cdot \hat{\mathbfcal{J}}\right)^2\Big)\label{app:hlhole11}
\ee 
with $\beta_1=\alpha_1$ and $\beta_2=\alpha_2+\alpha_3$, according to Eq.~(\ref{eq13}).

 \textbf{(i)} For $\vk$ along $\bf z$, the scalar product $\big(\vk\cdot \hat{\mathbfcal{J}}\big)^2$ is equal to $k^2 \hat{\mathcal{J}}_z^2$; so, Eq.~(\ref{app:hlhole10}) readily  gives the  eigenvalues of $\hat{H}^{(\rm sph)}_\vk$ as
\begin{subeqnarray}\label{app:hlhole12_0}
\frac{\hbar^2 k^2}{2m_0}\left(\frac{3}{2}\cdot\frac{5}{2}\beta_1+\left(\frac{3}{2}\right)^2\beta_2\right)=\va_{v,k}^{(H)}\equiv\frac{\hbar^2k^2}{-2m_H}\slabel{app:hlhole12}\\
\frac{\hbar^2 k^2}{2m_0}\left(\frac{3}{2}\cdot\frac{5}{2}\beta_1+\left(\frac{1}{2}\right)^2\beta_2\right)=\va_{v,k}^{(L)}\equiv\frac{\hbar^2k^2}{-2m_L}\slabel{app:hlhole13}
\end{subeqnarray}
The heavy and light electron eigenstates are noted  as $|v,\vk\ran\otimes|\mathcal{J}=\pm3/2\ran_z$ and $|v,\vk\ran\otimes|\mathcal{J}=\pm1/2\ran_z$, by analogy with the fourfold spin-orbit eigenstates of atoms. Since valence electrons have negative effective masses, heavy and light valence electrons correspond to $m_H>m_L>0$. This is fulfilled for $\beta_2>0>5\beta_1+3\beta_2$, that is, $\zeta_1>\zeta_3>1$, as obtained for usual semiconductors, due to  coupling and energy differences of the $|c,\vk\ran$ and $|c,\vk;\mu\ran$ spatial levels in the conduction band.

 \textbf{(ii)} For arbitrary wave vector $\vk$ with Euler angles ($\theta_\vk,\varphi_\vk$) in the $(x,y,z)$ crystal frame, $\vk\cdot \hat{\mathbfcal{J}}$ is equal to
\bea
\vk\cdot \hat{\mathbfcal{J}}&=&k\Big(\sin\theta_\vk \cos\varphi_\vk\hat{\mathcal{J}}_x +\sin\theta_\vk \sin\varphi_\vk\hat{\mathcal{J}}_y+\cos\theta_\vk\hat{\mathcal{J}}_z  \Big)
\nonumber
\\
&\equiv& k \hat{\mathcal{J}}_\vk\label{app:hlhole14}
\eea
 The operator $\hat{\mathcal{J}}_\vk$ follows from $\hat{\mathcal{J}}_z$ through a rotation of the quantization axis from $\bf{z}$ to $\vk$.  The above definition gives its $4\times 4$ matrix representation in the $|\mathcal{J}\ran_z$ eigenstate basis, with $\mathcal{J}=(3/2,1/2,-1/2,-3/2)$, as
\be
\hat{\mathcal{J}}_\vk=\hbar\left(\begin{matrix}
\frac{3}{2}c_\vk & \frac{\sqrt{3}}{2}s^*_\vk  & 0 & 0 \\
\frac{\sqrt{3}}{2}s_\vk& \frac{1}{2}c_\vk & s^*_\vk & 0 \\
0 & s_\vk & -\frac{1}{2}c_\vk & \frac{\sqrt{3}}{2}s^*_\vk \\ 
0 & 0 & \frac{\sqrt{3}}{2}s_\vk & -\frac{3}{2}c_\vk
\end{matrix}\right)_{z}\label{app:hlhole15}
\ee
with $c_\vk=\cos\theta_\vk$ and $s_\vk=e^{i\varphi_\vk}\sin\theta_\vk$.

It is easy to check that the $\hat{\mathcal{J}}_\vk$  eigenvalues also are $(\pm3\hbar/2,\pm\hbar/2)$. We thus conclude that the  eigenenergies of the spherical Hamiltonian $\hat{H}^{(\rm sph)}_\vk$ do not depend on the $\vk$ direction: they just are the heavy and light electron energies given in Eq.~(\ref{app:hlhole12_0}). By contrast, the heavy and light electron eigenstates $|\mathcal{J}\ran_\vk$, which respectively correspond to $\mathcal{J}=\pm3/2$ and $\mathcal{J}=\pm1/2$, depend on the $\vk$ direction through its Euler angles $(\theta_\vk,\varphi_\vk)$. A brute-force calculation gives these  eigenstates\cite{footnote2}
\be\label{app:hlhole16}
\hat{\mathcal{J}}_\vk|\mathcal{J}\ran_\vk=\hbar\mathcal{J}|\mathcal{J}\ran_\vk
\ee
in the $|\mathcal{J}\ran_z$ basis, for $\hat{\mathcal{J}}_\vk$ defined in Eq.~(\ref{app:hlhole15}), as  
\bea
\label{app:hlhole17}
| \mathcal{J}=3\eta/2\ran_\vk=\sum_{\sigma=\pm1} d_{\sigma\eta,\eta;\vk}^2\Big(d_{\sigma\eta,\eta;\vk}|3\sigma/2\ran_z
\nonumber
\\
+\sqrt{3}\, d_{-\sigma\eta,\eta;\vk}|\sigma/2\ran_z\Big)
\eea
for the two heavy electron states $\mathcal{J}=3\eta/2$ with $\eta=\pm1$, while for the two light electron states, $\mathcal{J}=\eta /2$, they are given by
\bea
|\mathcal{J}=\eta/2\ran_\vk=\eta\sum_{\sigma=\pm1} \sigma d_{\sigma\eta,\eta;\vk}\bigg(\Big(1-3|d_{-\sigma\eta,\eta;\vk}|^2\Big)|\sigma/2\ran_z
\nonumber
\\
-\sqrt{3}\, d_{\sigma\eta,\eta;\vk}d^*_{-\sigma\eta,\eta;\vk}|3\sigma/2\ran_z\bigg)\hspace{1cm}\label{app:hlhole18}
\eea
The $d_{\sigma,\eta;\vk}$ coefficients depend on the Euler angles   as
\be
d_{1,\eta;\vk}=\cos \frac{\theta_\vk}{2}\qquad d_{-1,\eta;\vk}=\eta e^{i\eta\varphi_\vk}\sin \frac{\theta_\vk}{2}\label{app:hlhole19}
\ee
The phases of the $|\mathcal{J}\ran_\vk$ eigenstates have been chosen such that $|\mathcal{J}\ran_\vk=|\mathcal{J}\ran_z$ for $\vk$  along $\bf z$: indeed, $d_{1,\eta;\vk}=1$ and $d_{-1,\eta;\vk}=0$ when $\theta_\vk=0$.

The above equations allow us to obtain the valence electron eigenstates within the spherical approximation (Fig.~\ref{appCoulomb_HL:fig2}(b)), in the $|\mathcal{J}\ran_z$ basis as
\be\label{app:hlhole21_0}
|v,\vk\ran\otimes|\mathcal{J}\ran_\vk=\sum_{\mathcal{J}'=(\pm3/2,\pm1/2)}|v,\vk\ran\otimes|\mathcal{J}'\ran_z\, {}_z\lan \mathcal{J}'|\mathcal{J}\ran_\vk
\ee

 \subsection{Physical understanding of ${}_{\vk'}\lan \mathcal{J}'|\mathcal{J}\ran_\vk$}

In the following, it will be useful to know the scalar products of spin-orbit eigenstates for heavy and light valence electrons, namely
\be\label{app:hlhole21_1}
{}_{\vk'}\lan \mathcal{J}'|\mathcal{J}\ran_\vk=\sum_{\mathcal{J}''}{}_{\vk'}\lan \mathcal{J}'|\mathcal{J}''\ran_z\,{}_z\lan \mathcal{J}''|\mathcal{J}\ran_\vk
\ee

Using Eqs.~(\ref{app:hlhole17},\ref{app:hlhole18}), we can show that these scalar products take a compact form
\bea
{}_{\vk'}\lan 3\eta'/2|3\eta/2\ran_\vk\!\!&=&\!\! \Big(D_{\eta',\vk';\eta,\vk}\Big)^3 \nn   \\
{}_{\vk'}\lan \eta'/2|3\eta/2\ran_\vk\!\!&=&\!\!\Big(D_{\eta',\vk';\eta,\vk}\Big)^2   \Big(\sqrt{3}\,D_{-\eta',\vk';\eta,\vk}\Big)
\label{app:hlhole21_2}
\\
{}_{\vk'}\lan \eta'/2|\eta/2\ran_\vk\!\!&=& \!\!\eta'\eta\Big( D_{\eta',\vk';\eta,\vk}\Big) \Big(1{-}\big|\sqrt{3}\, D_{-\eta',\vk';\eta,\vk}\big|^2\Big)
\nonumber
\eea
with $D_{\eta',\vk';\eta,\vk}$ defined as
\be\label{app:hlhole21_3}
D_{\eta',\vk';\eta,\vk}=\sum_{\sigma=\pm1} d^*_{\sigma\eta',\eta';\vk'}d_{\sigma\eta,\eta;\vk}
\ee

\textbf{(i)} For $\vk'$  parallel to $\vk$, Eq.~(\ref{app:hlhole19}) gives $d_{1,\eta';\vk}=d_{1,\eta;\vk}$ whatever $(\eta,\eta')$, while $d_{-1,\eta';\vk}=d_{-1,\eta;\vk}$ when $\eta'=\eta$, but $d_{-1,\eta';\vk}=-d^*_{-1,\eta;\vk}$ when $\eta'=-\eta$.
This ultimately gives $D_{\eta',\vk;\eta,\vk}=\delta_{\eta',\eta}$. For such $(\vk',\vk)$, the scalar products of Eq.~(\ref{app:hlhole21_2}) reduce to
\begin{subeqnarray}\label{app:hlhole21_4}
\delta_{\eta',\eta}&=&{}_{\vk}\lan 3\eta'/2|3\eta/2\ran_\vk={}_{\vk}\lan \eta'/2|\eta/2\ran_\vk\\
0&=&{}_{\vk}\lan \eta'/2|3\eta/2\ran_\vk
\end{subeqnarray}

\textbf{(ii)} For $\vk'$  antiparallel to $\vk$, that is, for $\vk'$ with Euler angles $(\pi + \theta_\vk,\varphi_\vk)$, Eq.~(\ref{app:hlhole19}) gives 
\bea
d_{1,\eta;\vk'}=d_{1,\eta;-\vk}=\cos\!\frac{\theta_\vk{+}\pi}{2}=-\sin\!\frac{\theta_\vk}{2}\hspace{1.8cm}\\
d_{-1,\eta;\vk'}=d_{-1,\eta;-\vk}=\eta e^{i\eta\varphi_\vk}\sin\! \frac{\theta_\vk{+}\pi}{2}{=}\eta e^{i\eta\varphi_\vk}\cos\! \frac{\theta_\vk}{2}
\eea 
 For such $(\vk',\vk)$, the scalar products of Eq.~(\ref{app:hlhole21_2})  reduce to
\begin{subeqnarray}\label{hlhole21_40}
{}_{-\vk}\lan 3\eta'/2|3\eta/2\ran_\vk&=&(-\eta e^{i\eta\varphi_\vk})^3\delta_{\eta',-\eta}\\
{}_{-\vk}\lan \eta'/2|\eta/2\ran_\vk&=&\eta e^{i\eta\varphi_\vk}\delta_{\eta',-\eta} \\
{}_{-\vk}\lan \eta'/2|3\eta/2\ran_\vk&=&0
\end{subeqnarray}
This shows that while heavy and light  electron states remain orthogonal, the ${}_{-\vk}\lan \mathcal{J}'|\mathcal{J}\ran_\vk$ matrix in the heavy-electron subspace or in the light-electron subspace are completely off-diagonal.

\textbf{(iii)} For arbitrary $\vk$ and $\vk'$,  the angle $\theta_{\vk'\vk}$ between $\vk$ and $\vk'$, reads in terms of the $\vk$ and $\vk'$ Euler angles as
\be\label{app:hlhole21_6}
\cos\theta_{\vk'\vk}=\cos\theta_{\vk'}\cos\theta_\vk+\sin\theta_{\vk'}\sin\theta_\vk\cos(\varphi_{\vk'}-\varphi_\vk)
\ee 
By noting from Eqs.~(\ref{app:hlhole19},\ref{app:hlhole21_3}) that
\be\label{app:hlhole21_5}
D_{\eta,\vk';\eta,\vk}= \cos\frac{\theta_{\vk'}}{2}\cos\frac{\theta_\vk}{2}{+}\sin\frac{\theta_{\vk'}}{2}\sin\frac{\theta_\vk}{2}e^{i\eta(\varphi_\vk-\varphi_{\vk'})}
\ee
\be\label{app:hlhole21_50}
D_{-\eta,\vk';\eta,\vk}=\eta\left(\cos\frac{\theta_{\vk'}}{2}\sin\frac{\theta_\vk}{2}e^{i\eta\varphi_{\vk}}{-}\sin\frac{\theta_{\vk'}}{2}\cos\frac{\theta_\vk}{2} e^{i\eta\varphi_{\vk'}}\right)
\ee
we can check that the $D_{\eta',\vk';\eta,\vk}$  moduli reduce to
\begin{subeqnarray}
\label{app:hlhole21_7}
\left|D_{\eta,\vk';\eta,\vk}\right|^2&=&\frac{1+\cos\theta_{\vk'\vk}}{2}=\cos^2\frac{\theta_{\vk'\vk}}{2}\\
\left|D_{-\eta,\vk';\eta,\vk}\right|^2&=&\frac{1-\cos\theta_{\vk'\vk}}{2}=\sin^2\frac{\theta_{\vk'\vk}}{2}
\end{subeqnarray}

To better understand the scalar products ${}_{\vk'}\lan \mathcal{J}'|\mathcal{J}\ran_\vk$ given in Eq.~(\ref{app:hlhole21_2}), we can introduce the unitary matrix $\hat{A}_{\vk',\vk}$, which is identical to the rotation matrix from $\vk$ to $\vk'$  in quantum mechanics which is defined in terms of $\hat{J}_y$ as $ e^{-i \hat{J}_y \theta_{\vk'\vk}}$. It reads in the  $(|3/2\ran_z,|1/2\ran_z,|-1/2\ran_z,|-3/2\ran_z)$ basis as
\be\label{app:hlhole21_9}
{\small\left(\begin{matrix}
C_{\vk'\vk}^3 & \cdot & \cdot & \cdot \\
{-}\sqrt{3}C_{\vk'\vk}^2 S_{\vk'\vk} & C_{\vk'\vk}(1{-}3S_{\vk'\vk}^2) & \cdot & \cdot \\
\sqrt{3}C_{\vk'\vk}S_{\vk'\vk}^2& S_{\vk'\vk}(1{-}3C_{\vk'\vk}^2) & C_{\vk'\vk}(1{-}3S_{\vk'\vk}^2)   & \cdot \\ 
{-}S_{\vk'\vk}^3 & \sqrt{3}S^2_{\vk'\vk}C_{\vk'\vk}  & {-}\sqrt{3}C_{\vk'\vk}^2 S_{\vk'\vk}  & C^3_{\vk'\vk}
\end{matrix}\right)}
\ee
for $C_ {\vk'\vk}=
\cos(\theta_{\vk'\vk}/2)$ and $S_ {\vk'\vk}=
\sin(\theta_{\vk'\vk}/2)$. We note that its $\hat{A}_{\vk',\vk}$ matrix elements fulfill
\bea
{}_z\lan \mathcal{J}'|\hat{A}_{\vk',\vk}|\mathcal{J}\ran_z&=&(-1)^{\mathcal{J}'-\mathcal{J}}{}_z\lan \mathcal{J}|\hat{A}_{\vk',\vk}|\mathcal{J}'\ran_z \nn\\
&=& {}_z\lan -\mathcal{J}|\hat{A}_{\vk',\vk}|-\mathcal{J}'\ran_z
\eea
So, by using Eqs.~(\ref{app:hlhole21_2},\ref{app:hlhole21_7}), we find that the modules of the $\hat{A}_{\vk',\vk}$ matrix elements reduce to
\be\label{eq34}
\big|{}_z\lan \mathcal{J}'|\hat{A}_{\vk',\vk}|\mathcal{J}\ran_z\big|=\big|{}_{\vk'}\lan \mathcal{J}'|\mathcal{J}\ran_\vk\big|
\ee

Since from the closure relation for $|\mathcal{J}\ran_\vk$ eigenstates, 
\bea\label{app:hlhole21_10}
\sum_{\mathcal{J}'=(\pm3/2,\pm1/2)}\!\!\!\!\!\! |{}_{\vk'}\lan \mathcal{J}'|\mathcal{J}\ran_\vk|^2&{=}&\sum_{\mathcal{J}'} {}_{\vk}\lan \mathcal{J}|\mathcal{J}'\ran_{\vk'}\,\,{}_{\vk'}\lan \mathcal{J}'|\mathcal{J}\ran_\vk
\nonumber
\\
&{=}&1
\eea
as can also be checked by summing up the squared column  elements of   the $\hat{A}_{\vk',\vk}$ matrix, we deduce that $|{}_{\vk'}\lan \mathcal{J}'|\mathcal{J}\ran_\vk|^2$ can be physically understood as the probability for the $|\mathcal{J}\ran_\vk$ state to end in the $|\mathcal{J}'\ran_{\vk'}$ state.

\section{Coulomb scatterings between heavy and light electrons\label{sec2}}

\subsection{Without spin-orbit coupling }

 In the absence of spin-orbit interaction, the  kinetic part of the valence electron Hamiltonian reads
\be\label{app:hlhole22}
\sum_\vk\sum_{\mu=(x,y,z)}\sum_{s=\pm1/2}\va_{v,\vk} \hat{a}^\dag_{v,\vk;\mu,s}\hat{a}_{v,\vk;\mu,s}
\ee
where $\hat{a}^\dag_{v,\vk;\mu,s}$ creates a valence electron with wave vector $\vk$ and spin $s$ in one of the three  spatial states $\mu$, the wave function of this electron, $\lan\vr|v,\vk;\mu\ran$, being given in Eq.~(\ref{2}).

The repulsive Coulomb interaction between valence and conduction electrons reads in terms of their creation operators as
\bea\label{app:hlhole23}
\hat{V}_{cv}=\sum_{\vq\not=\bf0}v_\vq  \sum_{\vk_1\vk_2} \,\, \sum_{s_1s_2}  \,\,  \sum_{\mu_2=(x,y,z)}
\,\,\,\,\,\,\,\,\,\,\,\,\,\,\,\,\,\,\,\,\,\,\,\,\,\,\,\,\,\,\,\,\,\,\,\,\,\,\,\,\,\,\,\,\,\,\,\,\,\,\,\,\,\,\,
\\
 \hat{a}^\dag_{c,\vk_1+\vq;s_1}\hat{a}^\dag_{v,\vk_2-\vq;\mu_2,s_2}\hat{a}_{v,\vk_2;\mu_2,s_2}\hat{a}_{c,\vk_1;s_1}
 \nonumber
\eea
with $v_\vq\simeq 4\pi e^2/L^3 \epsilon_{sc}q^2$ in the small wave-vector transfer limit. Note that the  electrons keep their quantum  indices, $s$ or $(\mu,s)$, in this direct Coulomb process.

\subsection{With spin-orbit coupling }

 The problem is  more complicated when the spin-orbit interaction is included because the spin-orbit eigenstates  $|\mathcal{J}\ran_\vk$ depend on the $\vk$ direction. This fundamentally means that when the $\vk$ wave vector of the electron changes, as  in a Coulomb process, its eigenstates also change.

The  kinetic part of the valence electron Hamiltonian in the spherical approximation, $\hat{H}^{(\rm sph)}_\vk$, is diagonal in its eigenstate basis $|v,\vk\ran\otimes|\mathcal{J}\ran_\vk$,
\bea\label{app:hlhole24}
\sum_\vk\sum_{\mathcal{J}=\pm3/2}\va^{(H)}_{v,k} \hat{a}^\dag_{v,\vk;\mathcal{J}_\vk}\hat{a}_{v,\vk;\mathcal{J}_\vk}
\,\,\,\,\,\,\,\,\,\,\,\,\,\,\,\,\,\,\,\,\,\,\,\,\,\,\,\,\,\,\,\,\,\,\,\,\,\,\,\,\,\,\,\,\,\,\,\,\,\,\,\,\,\,
\\
+\sum_\vk\sum_{\mathcal{J}=\pm1/2}\va^{(L)}_{v,k} \hat{a}^\dag_{v,\vk;\mathcal{J}_\vk}\hat{a}_{v,\vk;\mathcal{J}_\vk}
\nonumber
\eea
where $\hat{a}^\dag_{v,\vk;\mathcal{J}_\vk}$ creates a valence electron with wave vector $\vk$ in the $\mathcal{J}=\pm3/2$ spin-orbit level for heavy electrons and in the $\mathcal{J}=\pm1/2$ level for light electrons, 
\be\label{app:hlhole25}
|v,\vk\ran\otimes |\mathcal{J}\ran_\vk=\hat{a}^\dag_{v,\vk;\mathcal{J}_\vk} |vac\ran
\ee
where $|vac\ran$ denotes the vacuum state.

By contrast, the Coulomb part is not  diagonal in the $|\mathcal{J}\ran_\vk$ basis: a heavy valence electron can scatter into a heavy-mass state but also into a light-mass state. To derive these scatterings in a secure way, let us go step by step from the operator  $\hat{a}^\dag_{v,\vk;\mu,s}$ that appears in Eq.~(\ref{app:hlhole23}), to the succession of operators  $\hat{a}^\dag_{v,\vk;\mathcal{L}_z,s_z}$,   $\hat{a}^\dag_{v,\vk;\mathcal{J}_z}$, and ultimately to $\hat{a}^\dag_{v,\vk;\mathcal{J}_\vk}$ that appears in Eq.~(\ref{app:hlhole24}).

From Eq.~(\ref{app:hlhole6_0}) that leads to
\be\label{app:hlhole26}
\hat{a}^\dag_{\pm1_z}=\frac{\mp i\hat{a}^\dag_{x}+\hat{a}^\dag_{y}}{\sqrt{2}}\qquad\qquad
\hat{a}^\dag_{0_z}=i\hat{a}^\dag_{z}
\ee
we can check that 
\be\label{app:hlhole27}
\sum_{\mathcal{L}=(\pm1,0)}\hat{a}^\dag_{\mathcal{L}_z}\hat{a}_{\mathcal{L}_z}=\sum_{\mu=(x,y,z)} \hat{a}^\dag_\mu\hat{a}_\mu
\ee

Next, we  introduce the spin and use Eq.~(\ref{5}) to write the creation operators for  spin-orbit eigenstates as 
\begin{subeqnarray}\label{app:hlhole28}
\hat{a}^\dag_{\pm\frac{3}{2}_z}&=&\hat{a}^\dag_{\pm1_z;\pm\frac{1}{2}_z}\\
\hat{a}^\dag_{\pm\frac{1}{2}_z}&=&\frac{1}{\sqrt{3}}\left( \hat{a}^\dag_{\pm1_z;\mp\frac{1}{2}_z}+\sqrt{2}\,\hat{a}^\dag_{0_z;\pm\frac{1}{2}_z}\right)\\
\hat{a}'^\dag_{\pm\frac{1}{2}_z}&=&\pm\frac{1}{\sqrt{3}}\left(\sqrt{2}\, \hat{a}^\dag_{\pm1_z;\mp\frac{1}{2}_z}-\hat{a}^\dag_{0_z;\pm\frac{1}{2}_z}\right)
\end{subeqnarray}
These relations fulfill
\bea\label{app:hlhole29}
\sum_{\mathcal{J}=(\pm3/2,\pm1/2)} \hat{a}^\dag_{\mathcal{J}_z}\hat{a}_{\mathcal{J}_z}+\sum_{\mathcal{J}=\pm1/2} \hat{a}'^\dag_{\mathcal{J}_z}\hat{a}'_{\mathcal{J}_z}
\,\,\,\,\,\,\,\,\,\,\,\,\,\,\,\,\,\,\,\,\,\,\,\,\,\,\,\,\,\,\,\,\,\,\,\,\,
\\
\nonumber
=\sum_{\mathcal{L}=(\pm1,0)}\sum_{s=\pm1/2}\hat{a}^\dag_{\mathcal{L}_z,s_z}\hat{a}_{\mathcal{L}_z,s_z}
\eea

In the following, we will drop the split-off twofold subband with creation operators $\hat{a}'^\dag_{\mathcal{J}_z}$ and concentrate on the fourfold level with creation operators $\hat{a}^\dag_{\mathcal{J}_z}$ for $\mathcal{J}=(\pm3/2,\pm1/2)$ (see Fig.~\ref{appCoulomb_HL:fig2}(a)).

Using the above relation, we can rewrite the $\hat{V}_{cv}$ Coulomb interaction between conduction and valence electrons given in Eq.~(\ref{app:hlhole23})  as
\bea
\label{app:hlhole30}
\hat{V}_{cv}&=&\sum_{\vk_1\vk_2}\sum_{\vk'_2\not=\vk_2}v_{\vk'_2-\vk_2}
\sum_{s=\pm1/2}
\\
&{}&\sum_{\mathcal{J}=(\pm3/2,\pm1/2)}\!\! \hat{a}^\dag_{c,\vk_1+\vk_2-\vk'_2;s_z}\hat{a}^\dag_{v,\vk'_2;\mathcal{J}_z}\hat{a}_{v,\vk_2;\mathcal{J}_z}\hat{a}_{c,\vk_1;s_z}\nn
\eea
The above form is diagonal in the $|\mathcal{J}\ran_z$ spin-orbit basis, but these states do not have a well-defined kinetic energy within the spherical  Hamiltonian, $\hat{H}_\vk^{(\rm sph)}$.

By noting that
\bea\label{app:hlhole31}
\hat{a}^\dag_{v,\vk;\mathcal{J}_z}|vac\ran&=&|v,\vk\ran\otimes|\mathcal{J}\ran_z
\\
&=& |v,\vk\ran\otimes \!\!\!\sum_{\mathcal{J}'=(\pm3/2,\pm1/2)} \!\!\!|\mathcal{J}'\ran_\vk\,\,    {}_\vk\lan \mathcal{J}'|\mathcal{J}\ran_z\nn\\
&=& \sum_{\mathcal{J}'=(\pm3/2,\pm1/2)} \!\!\!{}_\vk \lan \mathcal{J}'|\mathcal{J}\ran_z \,\hat{a}^\dag_{v,\vk;\mathcal{J}'_\vk}|vac\ran
\nn
\eea
we can turn from $\hat{a}^\dag_{v,\vk;\mathcal{J}_z}$ to $\hat{a}^\dag_{v,\vk;\mathcal{J}_\vk}$, through 
\bea\label{app:hlhole32}
\sum_{\mathcal{J}}\hat{a}^\dag_{v,\vk';\mathcal{J}_z}\hat{a}_{v,\vk;\mathcal{J}_z}
\,\,\,\,\,\,\,\,\,\,\,\,\,\,\,\,\,\,\,\,\,\,\,\,\,\,\,\,\,\,\,\,\,\,\,\,\,\,\,\,\,\,\,\,\,\,\,\,\,\,\,\,\,\,\,\,\,\,\,\,\,\,\,\,\,\,\,\,\,\,\,\,\,\,\,\,\,\,\,\,\,\,\,\,\,\,\,\,\,\,
\\
=\sum_{\mathcal{J}'\mathcal{J}''}\hat{a}^\dag_{v,\vk';\mathcal{J}'_{\vk'}}\hat{a}_{v,\vk;\mathcal{J}''_{\vk}}
\sum_{\mathcal{J}}{}_{\vk'}\lan \mathcal{J}'|\mathcal{J}\ran_z \,\,{}_z \lan\mathcal{J}|\mathcal{J}''\ran_{\vk}
\nonumber
\eea
with the $\mathcal{J}$  sum reducing to the scalar product ${}_{\vk'}\lan\mathcal{J}'|\mathcal{J}''\ran_{\vk}$ given in Eq.~(\ref{app:hlhole21_2}). So, the $\hat{V}_{cv}$ interaction   ends by reading in terms of heavy and light valence electrons as
 \bea
  \label{app:hlhole33} 
 \hat{V}_{cv}&=&\sum_{\vk_1} \sum_{\vk_2}\sum_{\vk'_2\not=\vk_2}v_{\vk'_2-\vk_2}
 \\
   &{}&\sum _{s=\pm1/2}\,\,\,\,
 \sum_{(\mathcal{J},\mathcal{J}')=(\pm3/2,\pm1/2)}{}_{\vk'_2}\lan\mathcal{J}'|\mathcal{J}\ran_{\vk_2}
  \nn
  \\
 &{}&  \hat{a}^\dag_{c;\vk_1+\vk_2-\vk'_2;s_z}\hat{a}^\dag_{v,\vk'_2;\mathcal{J}'_{\vk'_2}}
\hat{a}_{v,\vk_2;\mathcal{J}_{\vk_2}}\hat{a}_{c;\vk_1;s_z}
 \nn 
\eea

A few points about this interaction are worth noting:

--- For $\vk'_2$ parallel to $\vk_2$, the scalar product $_{\vk'_2}\lan\mathcal{J}'|\mathcal{J}\ran_{\vk_2}$ is equal to $\delta_{\mathcal{J}',\mathcal{J}}$ according to Eq.~(\ref{app:hlhole21_4}). So, we only have transitions between the same heavy electrons or between the same light electrons. 

--- For $\vk'_2$ antiparallel to $\vk_2$, the scalar product $_{\vk'_2}\lan\mathcal{J}'|\mathcal{J}\ran_{\vk_2}$ is equal to $\delta_{\mathcal{J}',-\mathcal{J}}$ within a phase factor. So, we only have transitions between different heavy electrons or between different light electrons. 

--- For $\vk'_2$ not along $\vk_2$, transitions between the two types of heavy  electrons, between the two types of light  electrons, or between heavy and light electrons, are possible. For such wave vectors, the valence electron mass can change under a Coulomb scattering.

\begin{figure*}[t]
\begin{center}
\includegraphics[trim=0cm 6.5cm 0cm 6.5cm,clip,width=1.8\columnwidth]{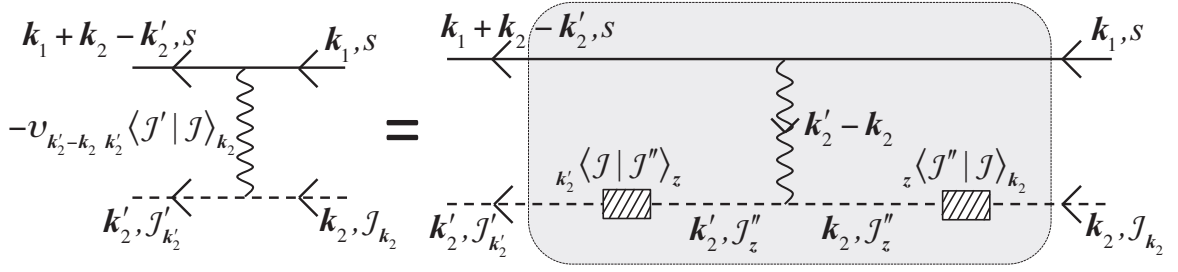}
\end{center}
\vspace{-0.4cm}
\caption{Effective Coulomb scattering between a conduction electron $(\vk_1,s)$ and a heavy or light valence hole $(\vk_2,\mathcal{J}_{\vk_2})$. The sum over the dummy index $\mathcal{J}''$ in ${}_{\vk'_2}\lan \mathcal{J}'|\mathcal{J}''\ran_z \,\,{}_z \lan\mathcal{J}''|\mathcal{J}\ran_{\vk_2}$, that has to be performed  in the grey box, leads to ${}_{\vk'_2}\lan\mathcal{J}'|\mathcal{J}\ran_{\vk_2}$. This scalar product shows up in the electron-hole scattering (diagram on the left) and renders this scattering not diagonal between heavy holes $\mathcal{J_\vk}=\pm3/2$ and light holes $\mathcal{J_\vk}=\pm1/2$.}
\label{appCoulomb_HL:fig1}
\end{figure*}

\section{From valence electrons to holes\label{sec3}}
The last step is to turn from valence electrons to holes. Because  tricky phase factors appear due to level degeneracy, let us repeat the procedure all over again.

\subsection{Procedure}

The change from valence electron destruction operator to hole creation operator  in a cubic crystal is based on two fundamental relations in the spatial  and spin subspaces, namely
\be\label{app:hlhole34}
\hat{a}_{v,\vk;\mu}=\hat{b}^\dag_{-\vk,\mu} \qquad \qquad \hat{a}_{\pm\frac{1}{2}}=\pm \hat{b}^\dag_{\mp\frac{1}{2}}
\ee
Equation (\ref{app:hlhole26}) then readily gives
\be\label{app:hlhole35}
\hat{a}_{\pm1_z}=\frac{\pm i\hat{a}_{x}+\hat{a}_{y}}{\sqrt{2}}=\hat{b}^\dag_{\mp1_z}\qquad\qquad
\hat{a}_{0_z}=-\hat{b}^\dag_{0_z}
\ee
When used into Eq.~(\ref{5}), we obtain
\be\label{app:hlhole36}
\hat{a}_{\pm\frac{3}{2}_z}=\hat{a}_{\pm1_z,\pm\frac{1}{2}_z}=\pm \hat{b}^\dag_{\mp1_z,\mp\frac{1}{2}_z}=\pm\hat{b}^\dag_{\mp\frac{3}{2}_z}
\ee
In the same way, 
\be\label{app:hlhole36_0}
\hat{a}_{\pm\frac{1}{2}_z}=\mp \hat{b}^\dag_{\mp\frac{1}{2}_z}\qquad \hat{a}'_{\pm\frac{1}{2}_z}=\mp \hat{b}'^\dag_{\mp\frac{1}{2}_z}
\ee

\subsection{Coulomb potential in terms of holes}

Since the Coulomb interaction between conduction and valence electrons, as given in Eq.~(\ref{app:hlhole30}), is diagonal in the $\mathcal{J}_z$ index, the phase factors that would appear in Eq.~(\ref{app:hlhole33}) cancel. As $\hat{b}_{-\vk'_2}\hat{b}^\dag_{-\vk_2}=-\hat{b}^\dag_{-\vk_2}\hat{b}_{-\vk'_2}$ since $\vk'_2\not=\vk_2$, we find that this interaction, when written in terms of electrons and holes, becomes\cite{footnote3}
 \bea
 \label{app:hlhole38}
  \hat{V}_{cv}&=&-\sum_{\vk_1\vk_2}\sum_{\vk'_2\not=\vk_2}v_{\vk'_2-\vk_2}
  \\
&{}&  \sum_{s=\pm1/2}\,\sum_{\mathcal{J}=(\pm3/2,\pm1/2)}\!\!\hat{a}^\dag_{\vk_1+\vk_2-\vk'_2,s}\hat{b}^\dag_{\vk'_2,\mathcal{J}_z}\hat{b}_{\vk_2,\mathcal{J}_z}\hat{a}_{\vk_1,s} 
\nonumber
 \eea

The last step is to  turn to heavy and light holes, following the procedure we have used for heavy and light valence electrons, that is, 
\be
\hat{b}^\dag_{\vk,\mathcal{J}_z}= \sum_{\mathcal{J}'} \, \hat{b}^\dag_{\vk,\mathcal{J}'_\vk}\,\,{}_\vk\lan \mathcal{J}'|\mathcal{J}\ran_z\label{app:hlhole39}
\ee
By summing over $\mathcal{J}$, we end with the Coulomb interaction between electron and heavy or light hole reading 
\bea
\label{app:hlhole40}
\hat{V}_{eh}&=& - \sum_{\vk_1 }\sum_{\vk_2} \sum_{\vk'_2\not=\vk_2}v_{\vk'_2-\vk_2}
\\
&{}&\sum_{s=\pm1/2}\,\,\sum_{(\mathcal{J},\mathcal{J}')=(\pm3/2,\pm1/2)}{}_{\vk'_2}\lan \mathcal{J}'|\mathcal{J}\ran_{\vk_2}\nonumber   \\
&&\times \hat{a}^\dag_{\vk_1+\vk_2-\vk'_2,s}\hat{b}^\dag_{\vk'_2,\mathcal{J}'_{\vk'_2}}\hat{b}_{\vk_2,\mathcal{J}_{\vk_2}}\hat{a}_{\vk_1,s}
\nonumber 
\eea

This interaction is visualized in the diagram of Fig.~\ref{appCoulomb_HL:fig1}. The minus sign evidences that it is attractive, as expected between electrons and holes, but not diagonal between heavy and light holes, except for processes in which $\vk'$ is parallel or antiparallel to $\vk$.

\section{Probabilities from heavy to light holes\label{sec4}}

The above equation (\ref{app:hlhole40}) shows that a heavy hole $\mathcal{J}_\vk=3\eta/2$ can keep its spin-orbit index or scatter into the other heavy hole state  $\mathcal{J}_\vk=-3\eta/2$. It can also scatter into one of the two light hole states  $\mathcal{J}_\vk=\pm1/2$. As mentioned  below Eq.~(\ref{app:hlhole21_10}), the squared modulus of the overlap ${}_{\vk'}\lan \mathcal{J}'|\mathcal{J}\ran_{\vk}$ can be physically understood as the probability to go from $|\mathcal{J}\ran_{\vk}$ to $|\mathcal{J}'\ran_{\vk'}$. The probability for the $\mathcal{J}_\vk=3\eta/2$  hole to stay heavy thus reads 
\bea
P_{HH}{=}\sum_{\eta'=\pm1}\big|{}_{\vk'}\lan 3\eta'/2|3\eta/2\ran_{\vk}\big|^2{=}\cos^6\frac{\theta_{\vk'\vk}}{2}+\sin^6\frac{\theta_{\vk'\vk}}{2}\nn\\
=1-\frac{3}{4}\sin^2\theta_{\vk'\vk}\hspace{4.9cm}
\eea 
with $\theta_{\vk'\vk}$ being the $(\vk',\vk)$ angle, while the probability for the $3\eta/2$ heavy hole to turn light is 
\bea
P_{LH}=\sum_{\eta'=\pm1}\big|{}_{\vk'}\lan \eta'/2|3\eta/2\ran_{\vk}\big|^2\hspace{3cm}
\nn\\
=3\cos^4\frac{\theta_{\vk'\vk}}{2}\sin^2\frac{\theta_{\vk'\vk}}{2}+3\sin^4\frac{\theta_{\vk'\vk}}{2}\cos^2\frac{\theta_{\vk'\vk}}{2}\nn\\
=\frac{3}{4}\sin^2\theta_{\vk'\vk}\hspace{4.9cm}
\eea

In the same way, the probability for the $\eta/2$ hole to stay light is
\bea
&&P_{LL}=\sum_{\eta'=\pm1}\big|{}_{\vk'}\lan \eta'/2|\eta/2\ran_{\vk}\big|^2\nn\\
&&=\cos^2\!\frac{\theta_{\vk'\vk}}{2}\Big(1{-}3\sin^2\!\frac{\theta_{\vk'\vk}}{2}\Big)^2{-}\sin^2\!\frac{\theta_{\vk'\vk}}{2}\Big(1{-}3\cos^2\!\frac{\theta_{\vk'\vk}}{2}\Big)^2\nn\\
&&=1-\frac{3}{4}\sin^2\theta_{\vk'\vk}
\eea 
while the probability for the $\eta/2$ light hole to turn heavy is given by
\be
P_{HL}=\sum_{\eta'=\pm1}\big|{}_{\vk'}\lan 3\eta'/2|\eta/2\ran_{\vk}\big|^2 =P_{LH}
\ee

 These probabilities  show that when $\vk'$ is parallel or antiparallel to $\vk$, the heavy holes stay heavy and the light holes stay light, in agreement with the results given in Eqs.~(\ref{app:hlhole21_4}) and (\ref{hlhole21_40}). For arbitrary $(\vk',\vk)$, we end with the remarkably simple result given in Eq.~(\ref{eq1}). Its precise derivation given here is very instructive, as it covers many tricky aspects of semiconductor physics.

\section{Conclusion}

In view of the notorious complexity of many-body effects in semiconductor physics, it has been common practice to start with an electron-hole Hamiltonian that is free from the crystal axes (spherical approximation), and to moreover take a single hole mass. These approximations are the ones that lead to  exciton looking very much like a hydrogen atom and that renders  many-body effects in semiconductors possible to handle.  

The existence of a hole mass difference has a dramatic consequence on Coulomb scatterings because these scatterings are not diagonal with respect to heavy and light holes. As a direct consequence, a heavy hole can turn light or vice versa, depending on the deflection angle in the scattering process at hand. In the present work, we give a detailed derivation of
the microscopic Coulomb scatterings between an electron and a heavy or light hole. This groundwork is the necessary first step toward exploring  new many-body effects resulting from the difference in heavy and light hole masses. We expect the Coulomb-mediated channel allowing heavy-light hole transition  to have an impact on the hole population distributions, the formation of mixed heavy-light-hole excitons, the dephasing in quantum beat phenomena\cite{Joschko,Ferrio}, to name a few.


\begin{thebibliography}{99}
\bibitem{footnote2} An alternative way is to derive the unitary operator $\hat{U}_\vk$ such that $\hat{\mathcal{J}}_\vk=\hat{U}_\vk^\dag \hat{\mathcal{J}}_z \hat{U}_\vk$. The $\hat{\mathcal{J}}_\vk$ eigenstates then follow from $\hat{U}_\vk^\dag$ acting on the $\hat{\mathcal{J}}_z$ eigenstates. However, this is neither more transparent nor simpler than the brute-force calculation.
\bibitem{footnote3} It is easier to first turn from electron to hole in the $\mathcal{J}_z$ basis, and then to change the hole basis from $\mathcal{J}_z$ to $\mathcal{J}_\vk$. The other way around is possible and gives the same result, but with more work.
\end{thebibliography}
\end{document}